\documentclass[12pt]{article}
\usepackage{epsfig}
\begin{document}
\vskip 2cm
\title{Soap Bubbles in Outer Space: Interaction of a Domain Wall with a
Black Hole}
\author{\\
M. Christensen${}^{*} {}^{1}$\, V.P. Frolov${}^{|} {}^{2,3,4}$\,
 and A.L. Larsen${}^{||} {}^{1}$}
\maketitle
\noindent
$^{1}${\em Institute of Physics, \ University of Odense, Campusvej 55, 5230
Odense M, Denmark.}\\
$^{2}${\em Theoretical Physics Institute, Department of Physics, University of
Alberta, Edmonton, Canada T6G 2J1.}
\\ $^{3}${\em CIAR Cosmology Program}
\\ $^{4}${\em P.N.Lebedev Physics Institute, \ Leninskii Prospect 53, \ \ Moscow
117924, \ Russia.}
\vskip 6cm
\noindent
$^{*}$Electronic address: mc@bose.fys.ou.dk\\
$^{|}$Electronic address: frolov@phys.ualberta.ca\\
$^{||}$Electronic address: all@fysik.ou.dk

\newpage
\begin{abstract}
\baselineskip=1.5em
We discuss the generalized Plateau problem in the 3+1 dimensional Schwarzschild
background. This represents the
physical situation, which could for instance have appeared in the early
universe,
where a cosmic membrane (thin domain wall) is located  near a black hole.
Considering stationary axially
symmetric membranes, three different membrane-topologies are possible
depending on the boundary conditions
at infinity:
2+1 Minkowski
topology, 2+1 wormhole  topology and 2+1 black hole topology.

Interestingly, we find that the different membrane-topologies are
connected via
phase transitions of the form first discussed by Choptuik in investigations
of scalar field collapse.
More precisely, we find a first order phase transition (finite mass gap)
between wormhole topology and
black hole topology; the intermediate membrane being an unstable wormhole
collapsing to a black hole.
Moreover, we find a second order phase transition (no mass gap) between
Minkowski topology and black hole
topology; the intermediate membrane being a naked singularity.

For the membranes of black hole topology, we find a mass scaling relation
analogous
to that originally found by Choptuik. However, in our case the parameter
$p$ is replaced by a
2-vector $\vec{p}$ parametrizing the solutions. We find that
$Mass\propto|\vec{p}-\vec{p}_*|^\gamma$
where $\gamma\approx 0.66$. We also find a periodic wiggle in the scaling
relation.

Our results show that  black hole formation
as a critical  phenomenon is far more general than expected.
\end{abstract}

\newpage
Cosmic strings and domain walls have played an
important role in theoretical cosmology and astrophysics (for a review of
topological defects, see
for instance \cite{vil}). Most of the work has
been devoted to cosmic strings, while domain walls have not attracted so
much attention.
In fact, it has been argued  that
stable domain walls are cosmologically disastrous. This was already pointed
out by
Zeldovich et. al. \cite{zel}, who considered domain wall structures
in models with spontaneous breaking of CP-symmetry. They argued that the
energy density of the domain walls is so large, that they would dominate
the universe completely,
violating the observed approximate isotropy and homogenity. So if domain
walls were ever formed in
the early universe, they were
 assumed to have somehow disappeared again, for instance by collapse,
evaporation or
simply by inflating away from our visible universe.

Much later however, Hill et. al. \cite{hill} introduced the so-called
"light"  domain walls. They considered a late-time
(post-decoupling) phase transition and found that light
domain walls could be produced, that were not necessarily in contradiction
with the observed large-scale structure of the universe.

Domain walls are formed in phase transitions where a discrete symmetry is
broken. Already from this, one
can argue that it is difficult to believe that domain walls {\it should
not} have been formed sometime
during the early evolution of the universe, where a number of phase
transitions certainly took place.
It is also worth mentioning that
domain walls and other topological  defects are now commonly seen
experimentally in
various areas
of condensed matter physics (for a review,
see for instance \cite{kib}).\\

In the leading approximation, a domain wall is described by the Dirac-Nambu-Goto
action \cite{dirac}
\begin{equation}
S = \mu \int d^3 \zeta \sqrt{- \det G_{AB}}
\end{equation}
where the induced metric on the world-volume is
\begin{equation}
G_{AB} = g_{\mu \nu} x^{\mu}_{,A} x^{\nu}_{,B} \label{inducedmetric}
\end{equation}
and $\mu$ is the tension. Here $x^{\mu}$ ($\mu = 0,1,2,3$) denote spacetime
coordinates while $\zeta ^A$ ($A = 0,1,2$) denote coordinates on the world
volume. The assumption made here is that the dimensions of the domain walls
are much greater than their thickness.
The domain wall is thus approximated by a relativistic membrane, which in
turn is assumed to be described by the
action (1). Using this model, the world-volume of the  membrane is a
minimal 2+1-surface embedded in a curved 3+1 dimensional
spacetime with metric $g_{\mu\nu}$. We are thus dealing with a
generalization of the classical Plateau problem in 3 Euclidean
dimensions (see for instance \cite{plateau}).

In this paper, we shall be interested in stationary axially symmetric membranes
embedded in the background of a Schwarzschild black hole. Using the
reparametrization invariances on the world-volume, a stationary axially
symmetric membrane can be parametrized by
\begin{equation}
t = \tau \; \; \; , \; \; \; r = \lambda \; \; \; , \; \; \;
\phi = \sigma \; \; \; , \; \; \; \theta = \theta (\lambda ) \label{gauge}
\end{equation}
where $(\tau , \sigma , \lambda )$ are the three coordinates on the
world-volume, and we use standard Schwarzschild
coordinates in target space. Then the action (1) reduces to
\begin{equation}
S_{{\mbox {eff}}} = 2 \pi \mu \Delta t \int r \, dr \, \sin \theta \,
\sqrt{1+r^2 (\theta ^{'} ) ^2 \left( 1-\frac{2 M}{r} \right)}
\end{equation}
where a prime denotes derivative with respect to $r$ $(=\lambda)$.
The corresponding equation of motion determining $\theta (r)$ is
\begin{equation}
\theta ^{''} + (2r-3M)(\theta ^{'})^3
- \frac{1}{\tan \theta} (\theta ^{'})^2
+ \frac{3r-4M}{r(r-2M)} \theta ^{'}
- \frac{1}{r(r-2M)\tan \theta} = 0 \label{eqnmotion}
\end{equation}
The line element on the world-volume, as obtained from (\ref{inducedmetric}),
is given by
\begin{equation}
d\Sigma ^2 = - \left( 1 - \frac{2M}{r} \right) dt^2 + \left( r^2
(\theta ^{'})^2
+ \left( 1 - \frac{2M}{r} \right)^{-1} \right) dr^2
+r^2 \sin ^2 \theta \;d \phi ^2 \label{lineelem}
\end{equation}
with $\theta (r)$ determined by (\ref{eqnmotion}),
while the  scalar curvature of the world-volume becomes
\begin{equation}
^{(3)}R = - 2\, \frac{\left( \theta ' \left( 1 - \frac{3M}{2r} \right)
- \frac{1}{r \tan  \theta  } \right) ^2
+ \left( \frac{\sqrt{3}M}{2r} \theta ' \right) ^2}{1 + r^2 (\theta ')^2
\left( 1 -  \frac{2M}{r} \right)} \label{scalcurv}
\end{equation}
In the following, we consider the membrane world-volume as a 2+1 dimensional
spacetime embedded in the background of the 3+1 dimensional Schwarzschild
spacetime.

In flat Minkowski spacetime ($M=0$), equation (\ref{eqnmotion}) is solved by
\begin{equation}
Z = \pm a \, \mbox{Arccosh} \left( \frac{R}{a}\right) + b \label{minmem}
\end{equation}
where $(a,b)$ are arbitrary constants ($a \geq 0$) and
\begin{equation}
(R,Z) = (r \sin \theta , -r \cos \theta)
\end{equation}
are the standard cylinder coordinates chosen such that the north pole
corresponds to $\theta=\pi$.

In 3 dimensional Euclidean geometry, the solution (8) is called a catenoid
\cite{plateau}.
In the relativistic setting here, it is more properly described as a 2+1
dimensional wormhole.
More precisely,
the corresponding world-volume line-element (\ref{lineelem}) is
\begin{equation}
d\Sigma ^2 = - dt^2 + \frac{R^2 }{R^2 - a^2} dR ^2 + R^2 d\phi ^2
\end{equation}
Thus, for $a \neq 0$ the membrane world-volume is a 2+1 dimensional wormhole,
while for $a = 0$ it is  2+1 dimensional Minkowski space.

We now consider equation (\ref{eqnmotion}) in the general case $M \neq 0$. We
have solved it numerically using the fourth order Runge-Kutta method. It is
convenient to introduce proper cylinder coordinates
\begin{equation}
(R_p,Z_p) = (l \sin \theta , -l \cos \theta)
\end{equation}
where $l$ is the proper radial distance
\begin{equation}
l = 2M + \sqrt{r(r-2M)} + 2M \ln \left(
\sqrt{\frac{r-2M}{2M}} + \sqrt{\frac{r}{2M}} \right)
\end{equation}
In these coordinates, the horizon of the 3+1 dimensional black hole corresponds
to
\begin{equation}
l_h = \sqrt{R_p^2 + Z_p^2} = 2M
\end{equation}

For the numerical integrations, we impose the following sets of boundary
conditions:\\
\begin{eqnarray}
&\mbox{I)}&\hspace*{1cm}R_p = 0 \; \; \; , \; \; \; Z_p > 2M \; \; \; , \;
\; \;
\frac{dZ_p}{dR_p} = 0 \label{bcI}\\
&\mbox{II)}&\hspace*{1cm}R_p > 2M \; \; \; , \; \; \; Z_p = 0 \; \; \; , \;
\; \;
\frac{dR_p}{dZ_p} = 0 \label{bcII}\\
&\mbox{III)}&\hspace*{1cm}\sqrt{R_p^2 + Z_p^2} = 2M \; \; \; , \; \; \;
\frac{d \theta}{dl} = 0 \label{bcIII}
\end{eqnarray}
These will describe all types of stationary axially symmetric and
$Z_2$-symmetric
(with respect to the equatorial plane) membranes in the Schwarzschild
background. Some
examples of solutions are shown in Fig.1.

The first set of solutions (I)
describe membranes which are always outside the 3+1 black hole. The
boundary conditions (14) are chosen to
ensure axial symmetry. The corresponding membranes  have the topology of
2+1 Minkowski space, and are deformed
versions of the $a=0$ membranes (\ref{minmem}) in flat 3+1 Minkowski space.

The second set of solutions (II) are also always outside the 3+1 black hole.
The boundary conditions (15) are chosen to
ensure $Z_2$-symmetry with respect to the equatorial plane of the 3+1
dimensional black hole.
The corresponding membranes have the topology of a 2+1 wormhole, and are
deformed versions of the
$b=0$ membranes (\ref{minmem}) in flat 3+1 Minkowski space. For these 2+1
wormholes, the 3+1 black hole is located in the middle of the throat in the
embedding diagram.

The third set of solutions (III) describe membranes entering the 3+1 black
hole. Notice that the boundary conditions (16) were chosen such that the
membranes cross the horizon of the 3+1
black hole orthogonally. This condition actually follows directly from the
equation of motion (5). More precisely, assuming
that $\theta'$ is regular at the horizon, the equation of motion (5) gives
the boundary condition
\begin{equation}
\theta'=\frac{1}{2M\tan\theta}\hspace*{1cm},\hspace*{1cm}r=2M
\end{equation}
Using $(R,Z)$-coordinates (9), this condition is equivalent to
$dZ/dR=0$, while in proper
cylinder coordinates (11) it becomes $d\theta/dl=0$. Thus in flat space
cylinder coordinates, the membranes cross the
horizon {\it horizontally} while in proper
cylinder coordinates, they cross the horizon {\it orthogonally}. The
picture is thus analogue to that of magnetic field
lines crossing the 3+1 black hole horizon \cite{thorn}. Obviously this
third family of membranes entering the 3+1 black hole
have no counterpart in flat 3+1 Minkowski space. These
membranes  have themselves the topology of a 2+1
dimensional black hole, as follows from equations (\ref{lineelem}) and
(\ref{scalcurv}): They have a spacetime singularity at $r \sin \theta = 0$
hidden behind the horizon located at $r_h = 2M$.

Each solution from any of the three families of membranes is uniquely
specified by its
asymptotic behaviour.
Asymptotically, the solution to equation (\ref{eqnmotion}) is of the same
form as (\ref{minmem}):
\begin{equation}
Z_p = \pm a_p \, \mbox{Arccosh} \left( \frac{R_p}{a_p}\right) + b_p
\end{equation}
where $(a_p,b_p)$ are constants ($a_p \geq 0$). That is, a membrane is
specified by a 2-vector
\begin{equation}
\vec{p}=\left(\begin{array}{c}
a_p\\ b_p\end{array}\right)
\end{equation}
Numerically we can then compute $\vec{p}=(a_p,b_p)$ corresponding to the 3
sets of
boundary
conditions (\ref{bcI})-(\ref{bcIII}). The result is shown in Fig.2. In this
plot, each point $\vec{p}=(a_p,b_p)$ corresponds to a stationary axially
symmetric and
$Z_2$-symmetric membrane embedded in the 3+1 dimensional Schwarzschild
background. Thus, the two components of $\vec{p}$ are not independent.

We
are particularly interested in the "phase transitions" between the different
membrane topologies, as discussed above. At the point $\vec{p}_0 = (0,0)$ there
is a transition between wormhole topology and black hole topology. The limiting
membrane is an "unstable" wormhole which collapses in the Z-direction and
becomes what is formally a 2+1 dimensional Schwarzschild spacetime:
\begin{equation}
d \Sigma ^2 = - \left( 1- \frac{2M}{r} \right) dt^2 +
\left( 1- \frac{2M}{r} \right) ^{-1} dr^2 + r^2 d \phi ^2
\end{equation}
with horizon at $r_h = 2M$ and singularity at $r=0$. This membrane of
course just corresponds to the
equatorial plane of the 3+1 dimensional Schwarzschild black hole.

The other transition happens at the point
\begin{eqnarray}
\vec{p}_* = (0.3048...,
2.0457...)\nonumber
\end{eqnarray}
 This is a
transition between Minkowski topology and black hole topology. It is an
interesting observation that the transition point in parameter space is
approached by infinite logaritmic spirals from both sides. This can be seen
by doing a conformal
transformation that blows up the region near $\vec{p}_*$
\begin{equation}
\vec{p}\longrightarrow \vec{p}\;'=\left(\begin{array}{c}
a'_p\\ b'_p\end{array}\right)=\frac{\vec{p}-\vec{p}_*}{|\vec{p}-\vec{p}_*|\;
|\ln |\vec{p}-\vec{p}_*||}
\end{equation}
The result of this transformation is  shown in Fig.3. Moreover, the
limiting membrane corresponding to $\vec{p}_*$ is a 2+1 dimensional naked
singularity as follows
from equation (\ref{scalcurv}): The singularity of this membrane is located
at $(r,\theta ) =
(2M,\pi )$, and it is not hidden behind a horizon.\\

These results are very similar to analogue results, first obtained by Choptuik
\cite{chop}, for the spherical collapse of scalar or Yang-Mills fields in 3+1
dimensions (for a review, see \cite{gund}).

In particular, in the case of Yang-Mills collapse \cite{chop2} or massive scalar
field collapse \cite{brady}, two different types of phase transitions occur at
the threshold of black hole formation: A first order phase transition (finite
mass gap) where the limiting solution is an unstable soliton star and a second
order phase transition (no mass gap) where the limiting solution is a naked
singularity.

For our membranes of 2+1 dimensional black hole topology, the mass inside the
apparent horizon $S$ can be defined by (up to normalization) \cite{wald}
\begin{equation}
Mass = - \frac{1}{4 \pi} \int _S \varepsilon _{ABC}  \nabla ^B \xi ^C d
\zeta ^A
\end{equation}
where $\nabla _B$ is the covariant derivative with respect to the metric
(\ref{inducedmetric}), and $\xi ^C$ is the timelike-at-infinity Killing vector
on the world-volume.
Using (\ref{inducedmetric}) and (\ref{gauge}) we get:
\begin{equation}
Mass = \frac{\sin \theta _0}{2}
\end{equation}
where $\theta _0$ is the polar angle at which the membrane crosses the horizon
of the 3+1 dimensional black hole. It should be stressed that (23) is the
mass inside the apparent
horizon of the membrane world-volume.
Since the membrane world-volume (6) is not a  vacuum solution in 2+1
dimensions, the mass (23) does not
equal the mass measured at infinity (for a discussion of the different mass
definitions, see
for instance
\cite{wald}).  Notice also that  our units and conventions are
such that the mass (23) is dimensionless. This corresponds to units where
the 3-dimensional gravitational
constant equals unity.

From (23) follows that the transition between
wormhole topology and black hole topology ($\theta _0 = \frac{\pi}{2}$) is of
first order (finite mass gap) while the transition between 2+1 Minkowski
topology and black hole topology ($\theta _0 = \pi$) is of second order (no mass
gap).

A generic result  of the investigations of scalar field collapse
\cite{chop} (see \cite{gund}
for a review) is a mass scaling relation of the form
$M_{BH}\propto|p-p_*|^\gamma$, where $p$
parametrizes the solutions  and $p_*$ is the critical parameter defined
such that a black hole
is formed for $p>p_*$. In our case of stationary membranes, the parameter $p$ is
replaced by the 2-vector $\vec{p}$, c.f. eqs.(18)-(19), while the mass of
the black hole is given
by (23).

In Fig.4., we show a double-logaritmic plot of $(Mass)$ versus
$|\vec{p}-\vec{p}_*|$. It corresponds to a relation of the form
\begin{equation}
\ln(Mass)=\gamma\ln|\vec{p}-\vec{p}_*|\;\;+\;\;\mbox{periodic function}
\end{equation}
where an additive constant has been absorbed in the periodic function.
This is a mass scaling relation analogues to that of Choptuik, including
the periodic wiggle with
period $\omega$ \cite{gund2, hod, garf} ($\omega$ is the period in $\ln
|\vec{p}-\vec{p}_*|\;$).
Numerically we find the following values of the parameters
\begin{eqnarray}
\gamma&\approx &0.66\nonumber\\
\omega&\approx &3.56\nonumber
\end{eqnarray}
The periodic function reflects the periodic self-similarity of the critical
solution
\cite{gund2, hod, garf}, already present in the original investigation
\cite{chop}. It should also be
mentioned that more precise numerical computations indicate that
$\gamma=2/3$, but at the present moment
we have no analytical proof of this.\\

It is also interesting to compare with the case of stationary cosmic
strings in the background of a black
hole \cite{fro}. In this case, the world-sheet of the string can be
considered a 1+1 dimensional
spacetime. Depending on the boundary conditions at infinity, the topology
of the string world-sheet is
either that of 1+1 Minkowski spacetime or that of a 1+1 black hole
\cite{all, hen}. Also in this case,
there is a phase transition between the two topologies. However, this phase
transition is of first order,
that is to say, there is a finite mass gap. Thus in the stationary string
case, there is no phase
transition of second order and no mass scaling relation of the type
originally discovered by Choptuik.\\

In conclusion, using analytical and numerical methods, we have considered
the interaction of a domain
wall with a Schwarzschild black hole. As a result we
have shown that, although our physical setup is completely different, the
phenomena concerning black
hole formation in 2+1 dimensions are very similar to those observed for
gravitational
collapse of various fields in 3+1 dimensions.
This again confirms the generality of black hole formation as a critical
phenomenon \cite{chop}, involving different types of phase transitions.
And most importantly,  our results show that  black hole formation
as a critical  phenomenon is far more general than expected.

\newpage
\noindent
\centerline{\psfig{file=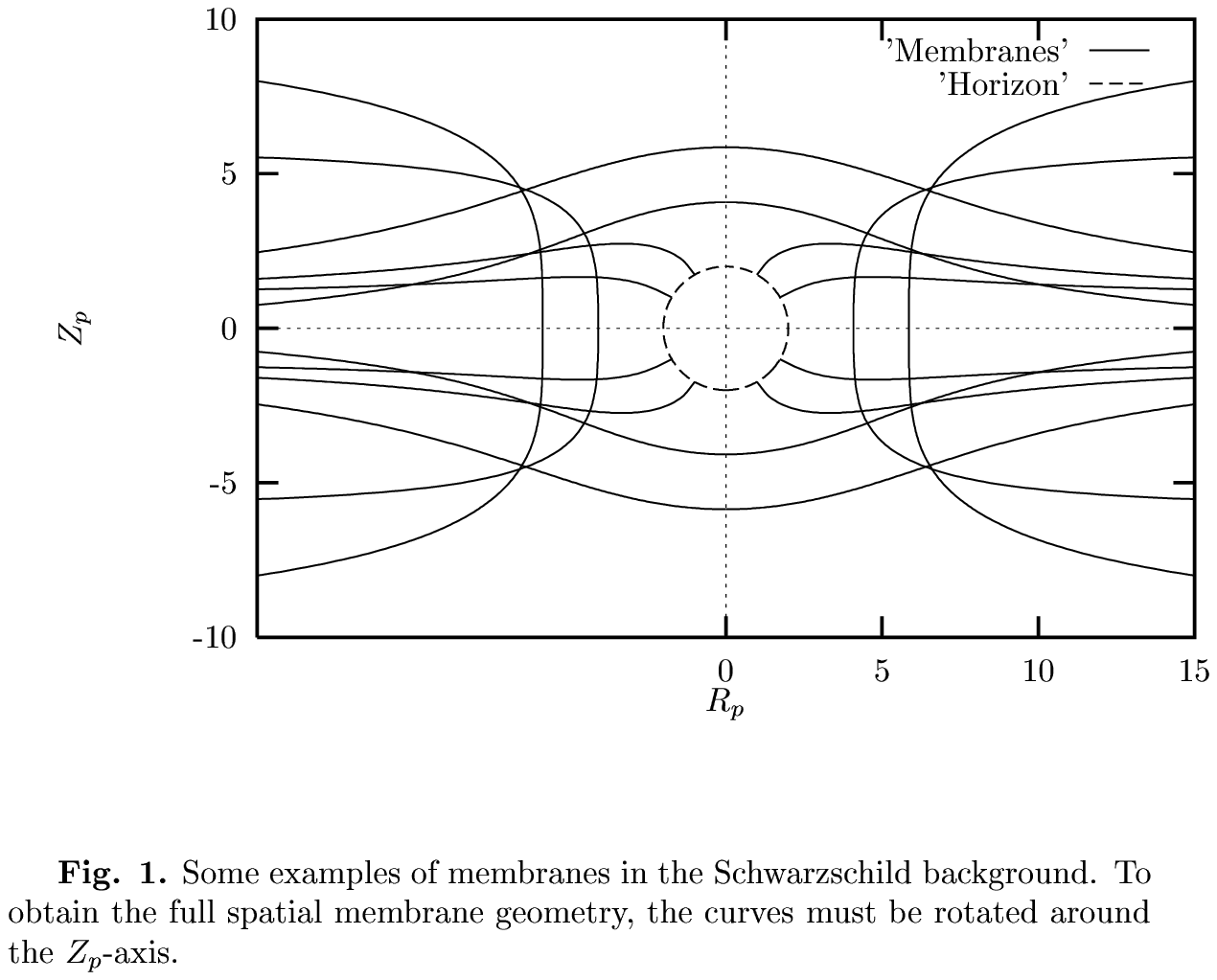,height=30cm,angle=0}}

\newpage
\noindent
\centerline{\psfig{file=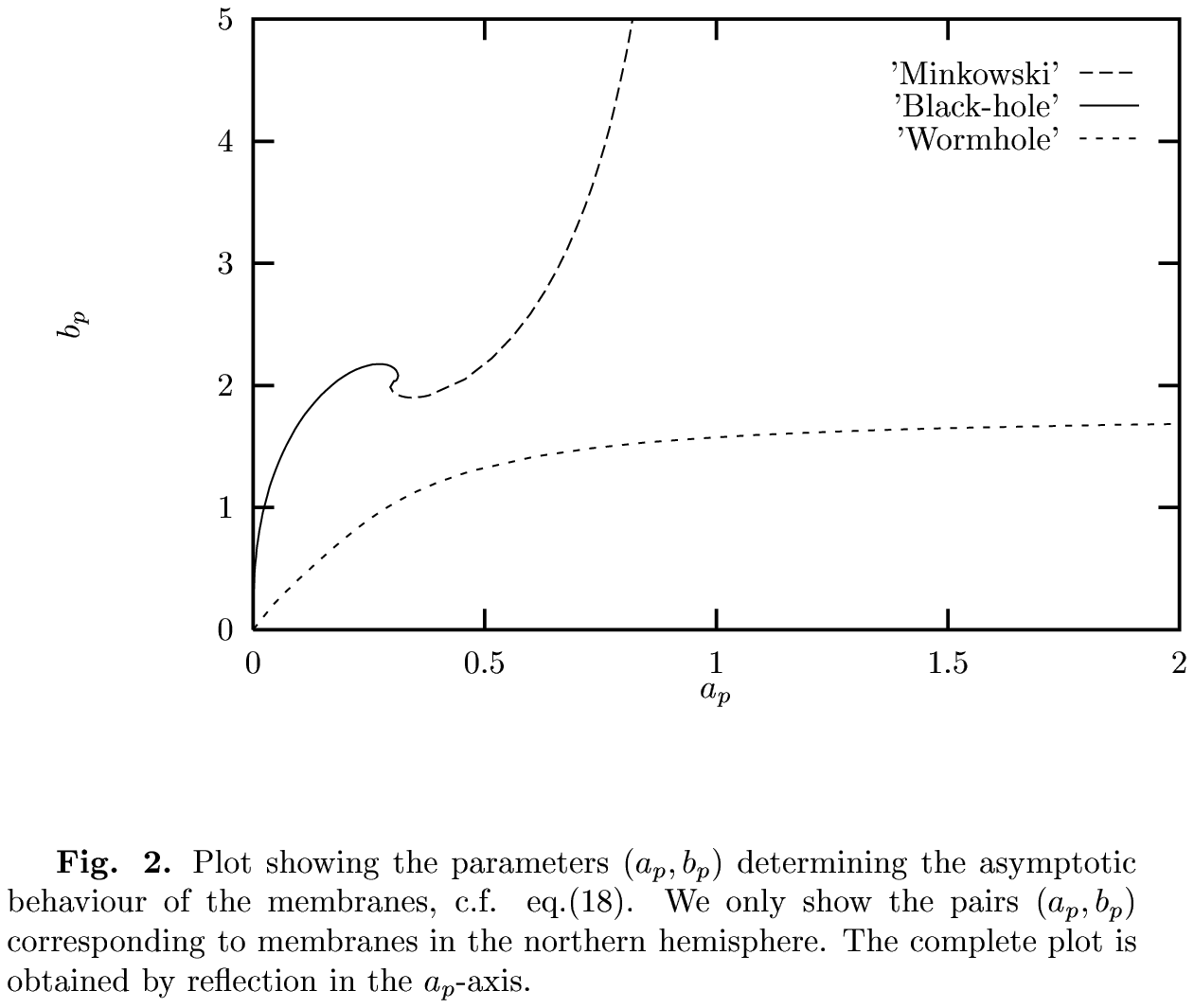,height=30cm,angle=0}}

\newpage
\noindent
\centerline{\psfig{file=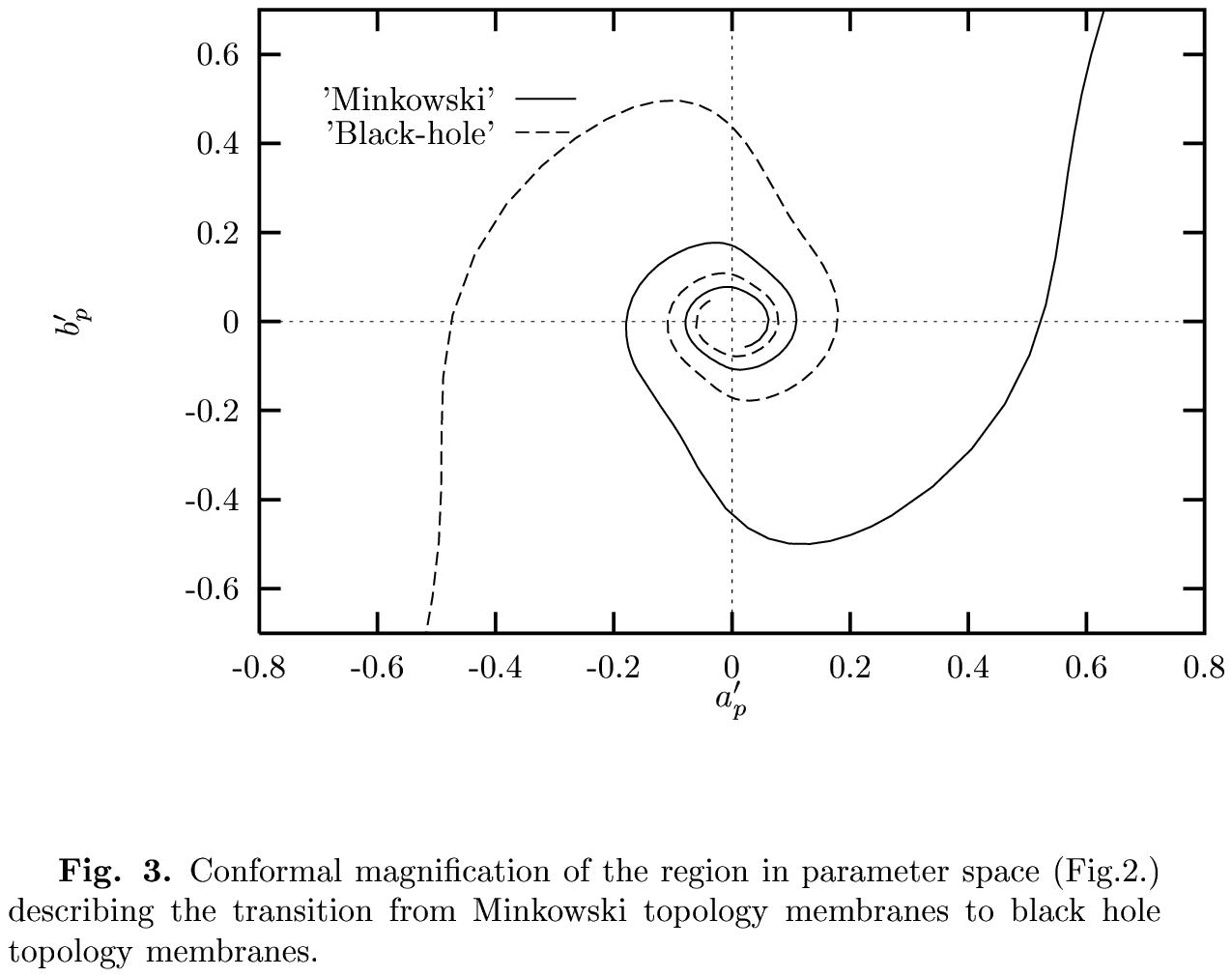,height=30cm,angle=0}}

\newpage
\noindent
\centerline{\psfig{file=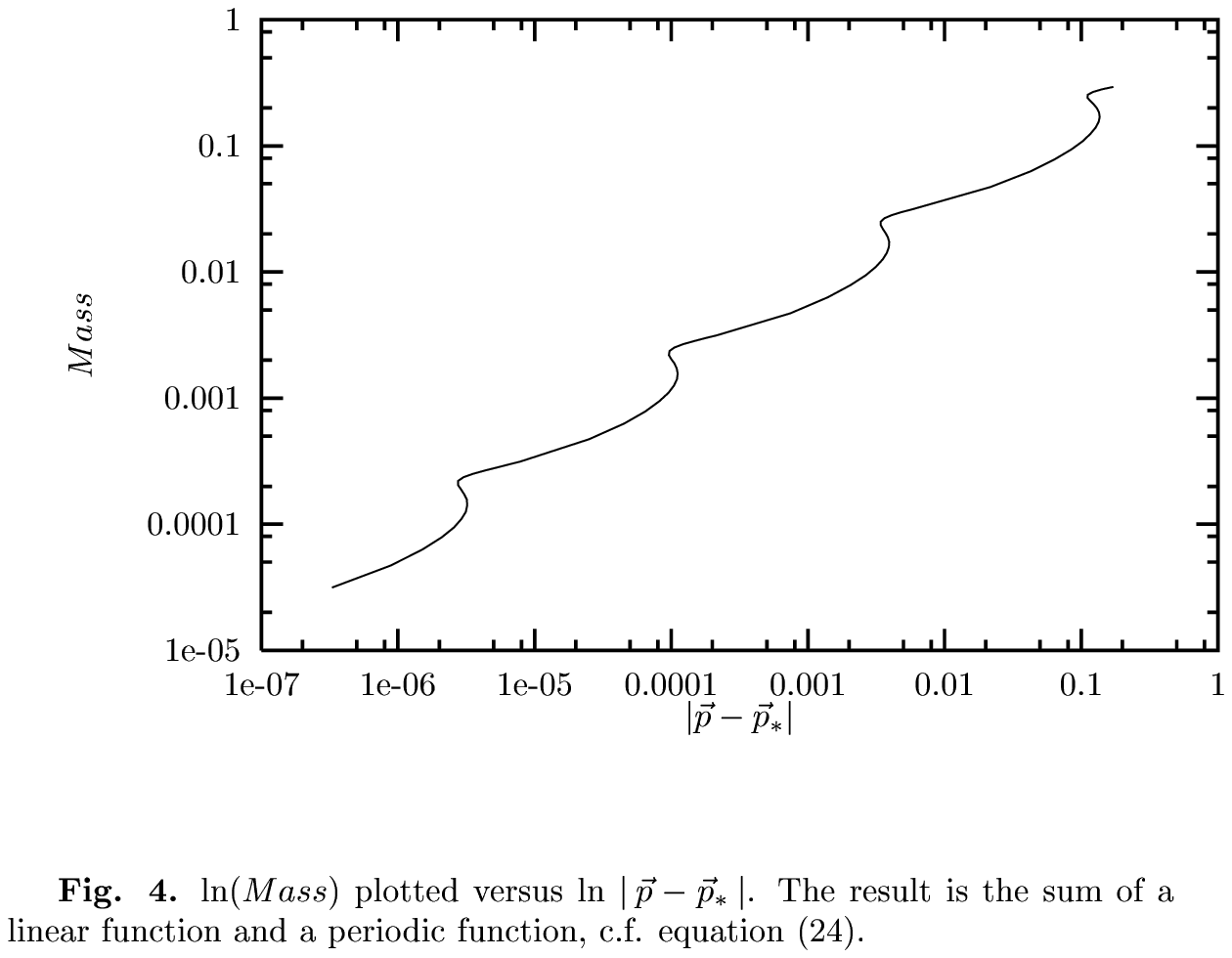,height=30cm,angle=0}}


\newpage
\begin{thebibliography}{11}
\bibitem{vil}A. Vilenkin and E.P.S. Shellard, {\it Cosmic Strings and other
Topological Defects}
(Cambridge University Press, Cambridge, 1994).
\bibitem{zel}Y.B. Zeldovich, I.Y. Kobzarev and L.B. Okun, Zh. Eksp. Teor.
Fiz {\bf 67} (1974) 3.
\bibitem{hill}C.T. Hill, D.N. Schramm and J.N. Fry, Comments Nucl. Part.
Phys. {\bf 19} (1989) 25.
\bibitem{kib}T.W.B. Kibble, in {\it Current Topics in Astrofundamental
Physics}, Proceedings of the International School of
Astrophysics "D. Chalonge" held in September 1996. Edited by N. Sanchez and
A. Zichichi  (World Scientific, Singapore,
1997). Pages 322-342.
\bibitem{dirac}P.A.M. Dirac, Proc. R. Soc. London, {\bf 268A} (1962) 57.
\bibitem{plateau}H.B. Lawson, Jr., {\it Lectures on Minimal Submanifolds}
Vol 1 (IMPA, Rio de Janeiro, 1973). Chapters 2, 3.
\bibitem{thorn}K.S. Thorne, R.H. Price and D.A. Macdonald, {\it Black
Holes: The Membrane Paradigm}
(Yale University Press, New Haven, 1986). Pages 24-25.
\bibitem{chop}M.W. Choptuik, Phys. Rev. Lett. {\bf 70} (1993) 9.
\bibitem{gund}C. Gundlach, "Critical phenomena in gravitational collapse",
Preprint, gr-qc/9712084.
\bibitem{chop2}M.W. Choptuik, T. Chmaj and P. Bizon, Phys. Rev. Lett. {\bf
77} (1996) 424.
\bibitem{brady}P.R. Brady, C.M. Chambers and S.M.C.V. Goncalves, Phys. Rev.
{\bf D56} (1997) R6057.
\bibitem{wald}R.M. Wald, {\it General Relativity} (The University of
Chicago Press, Chicago, 1984).
Chapter 11.2.
\bibitem{gund2}C. Gundlach, Phys. Rev. {\bf D55} (1997) 695.
\bibitem{hod}S. Hod and T. Piran, Phys. Rev. {\bf D55} (1997) 440.
\bibitem{garf}D. Garfinkle and G.C. Duncan, "Scaling of curvature in
sub-critical gravitational collapse",
Preprint, gr-qc/9802061.
\bibitem{fro}V.P. Frolov, V. Skarzhinski, A. Zelnikov and O. Heinrich,
Phys. Lett. {\bf B224} (1989) 255.
\bibitem{all}V.P. Frolov and A.L. Larsen, Nucl. Phys. {\bf B449} (1995) 149.
\bibitem{hen}V.P. Frolov, S. Hendy and A.L. Larsen, Phys. Rev. {\bf D54}
(1996) 5093.
\end{thebibliography}
\end{document}